\documentclass[12pt]{article}

\usepackage{epsfig}
\usepackage{latexsym}
\usepackage{amsfonts}
\usepackage{amsmath}
\usepackage{amsthm}
\usepackage{amssymb}
\usepackage{amsbsy}
\usepackage{slashed}
\usepackage{color}

\def\calm         {{\cal M}}

\newsavebox{\uuunit}
\sbox{\uuunit}
    {\setlength{\unitlength}{0.825em}
     \begin{picture}(0.6,0.7)
        \thinlines
        \put(0,0){\line(1,0){0.5}}
        \put(0.15,0){\line(0,1){0.7}}
        \put(0.35,0){\line(0,1){0.8}}
       \multiput(0.3,0.8)(-0.04,-0.02){12}{\rule{0.5pt}{0.5pt}}
     \end {picture}}

\def\be{\begin{equation}}
\def\ee{\end{equation}}
\def\bea{\begin{eqnarray}}
\def\eea{\end{eqnarray}}

\newcommand{\beq}{\begin{eqnarray}}
\newcommand{\eeq}{\end{eqnarray}}

\newcommand{\eal}[1]{\be \begin{aligned} #1 \end{aligned}\ee}

\newcommand{\ti}[1]{\tilde #1}

\def\sF{{{ F}\!\!\!\!\hskip.8pt\hbox{\raise1pt\hbox{/}}\,}}
\def\som{{{ \omega}\!\!\!\!\hskip.8pt\hbox{\raise1pt\hbox{/}}\,}}
\def\sJ{{{\rm J}\!\!\!\!\hskip.8pt\hbox{\raise1pt\hbox{/}}\,}}

\newcommand{\SL}{\mathop{\rm SL}}

\newcommand{\SLn}[2]{\SL({#1},\mathbb{#2})}

\renewcommand{\Im}{\operatorname{Im}}
\newcommand\e{\mathrm{e}}
\newcommand\iu{\operatorname{i}}
\newcommand\diff{\mathrm{d}}

\begin{document}
 
\begin{titlepage}

\begin{flushright}
{IPhT-T11/210}\\
\end{flushright}
\bigskip
\bigskip
\bigskip
\centering{\Large \bf Camouflaged Supersymmetry \vspace{0.9cm}}\\
\bigskip
\bigskip
\centerline{{\bf Iosif Bena, Hagen Triendl, Bert Vercnocke}}
\bigskip
\centering{Institut de Physique Th\'eorique, \\ CEA Saclay, 91191 Gif sur Yvette, France\vspace{0.4cm}}
\bigskip
\bigskip
\centerline{{ iosif.bena@cea.fr,~hagen.triendl@cea.fr,~bert.vercnocke@cea.fr} }
\bigskip
\bigskip

\begin{abstract}
 We establish a relation between certain classes of flux compactifications and certain families of black hole microstate solutions. This connection reveals a rather unexpected result: there exist {\it supersymmetric} solutions of $N=8$ supergravity that live inside many $N=2$ truncations, but are not supersymmetric inside any of them. If this phenomenon is generic, it indicates the possible existence of much larger families of supersymmetric black rings and black hole microstates than previously thought.

\end{abstract}
\end{titlepage}

\section{Introduction}

There is an extensive body of work on obtaining supersymmetric and non-supersymmetric vacua for flux compactifications of string theory and studying their phenomenology, and a parallel extensive body of work on constructing supersymmetric and non-supersymmetric black hole microstate solutions to understand black hole physics in string theory. While the physical motivations are different, the technical tools are rather close. 
In particular, the equations underlying supersymmetric solutions are well-understood and classified: On the flux compactification side (see for example \cite{Grana:2000jj,Giddings:2001yu,Grana:2004bg}) in ten dimensions, on the black hole microstate side for the underlying supergravity in five dimensions \cite{Gauntlett:2002nw,Gutowski:2004yv,Bena:2004de}.
Furthermore, some of the methods for constructing non-supersymmetric solutions from supersymmetric ones are strikingly similar. These methods include slightly deforming the supersymmetric solution by additional fluxes \cite{Grana:2000jj,Giddings:2001yu}, flipping some signs \cite{Goldstein:2008fq}, or writing some effective Lagrangian as a sum of squares for black holes \cite{Ferrara:1997tw,Denef:2000nb,Ceresole:2007wx,Andrianopoli:2007gt,LopesCardoso:2007ky} or flux backgrounds \cite{Lust:2008zd,Held:2010az}.

It is therefore not surprising that one can find a relation between certain types of solutions on the two sides. Indeed, as we will show below, certain  supersymmetric flux backgrounds of the type \cite{Becker:1996gj} where the ``internal'' (non-compact) manifold contains a hyper-K\"ahler factor can be interpreted as certain non-rotating solutions in the classification of \cite{Gauntlett:2002nw,Gutowski:2004yv,Bena:2004de}. (One can similarly relate non-supersymmetric solutions. The story is more intriguing and will be alluded to in this letter, but we leave the details for a companion publication \cite{Bena:2011aa}.) The main purpose of this letter is to show that there  are other supersymmetric solutions of the same class of flux compactifications which, when interpreted as black hole microstates in $N=2$ supergravity, do {\em not} fall into the classification of supersymmetric solutions \cite{Gauntlett:2002nw,Gutowski:2004yv,Bena:2004de}.\footnote{We use four-dimensional supersymmetry conventions. For instance, all $N=2$ theories, regardless of dimension, have 8 supercharges.} Hence, from the point of view of $N=2$ supergravity, these solutions should be non-supersymmetric. However, they are supersymmetric inside $N=8$ supergravity!

As we will explain below, these solutions have the right field content to fit into many possible $N=2$ truncations, and hence they will always be solutions of these $N=2$ theories.  However, the unbroken supercharges are projected out in all possible $N=2$ truncations and hence from the point of view of $N=2$ supergravity none of these solutions are supersymmetric\footnote{The fact that a non-supersymmetric solution of an $N=2$ or an $N=4$ theory can become supersymmetric when embedded in $N=8$ has been know for quite a while \cite{Ferrara:2007pc, Khuri:1995xq}. However, in all these examples, there always exists an $N=2$ or $N=4$ truncation in which the $N=8$ solution is supersymmetric. In our example no such truncation exists, and the supersymmetry of the solution cannot be captured in any daughter $N=2$ theory.}. A simple way to understand this is to recall that all $N=2$ supersymmetric solutions in the class \cite{Gauntlett:2002nw,Gutowski:2004yv,Bena:2004de} have (in our conventions) anti-self-dual fields on a hyper-K\"ahler base, while our solutions have {\it both} anti-self-dual and self-dual fields.

Our results have quite a few unexpected implications. First, it is widely believed that all supersymmetric microstate geometries of three-charge black holes in five dimensions are described by the equations of \cite{Gauntlett:2002nw,Gutowski:2004yv,Bena:2004de}. Our results indicate that many solutions that are not described by these equations are also supersymmetric in the parent $N=8$ theory. This implies that beside the classes of microstate solutions constructed so far there may exist many more supersymmetric microstates, which would contribute to the entropy count.

Second, it has been conjectured \cite{Ferrara:2006yb} and argued \cite{Bossard:2010mv} that all multicenter supersymmetric solutions of $N=8$ supergravity must live inside an $N=2$ truncation, and one may believe that this implies that the solutions of \cite{Denef:2000nb,Bates:2003vx} capture all supersymmetric multicenter $N=8$ solutions. Our results show that this is not so. 

Third, it is well-known that the supersymmetric black ring in five dimensions \cite{Elvang:2004rt,Bena:2004de,Elvang:2004ds, Gauntlett:2004qy} is part of a truncation to $N=2$ ungauged supergravity and belongs to the class of solutions \cite{Gauntlett:2002nw,Gutowski:2004yv,Bena:2004de}. Our results indicate that there may exist a new, more general supersymmetric black ring with more dipole charges (coming from the extra self-dual fluxes). Besides its interest as a new solution, if this black ring existed, it may also help to account for the missing entropy between the D1-D5 CFT and the dual bulk in the moulting black hole phase \cite{Bena:2011zw}.

More generally, the relation between black hole microstates and flux compactifications that we outline will likely prove fruitful in both directions.
There exists a whole methodology for constructing flux compactifications by writing the effective Lagrangian governing these compactifications as a sum of squares of calibrations \cite{Lust:2008zd,Held:2010az}. Under the guise of ``floating branes'', calibrations have also been used to find non-supersymmetric black hole microstates \cite{Bena:2009fi}, and relating the two approaches is likely to yield novel classes of solutions on both sides. We plan to report on this relation in an upcoming companion paper \cite{Bena:2011aa}. 
Furthermore, it has been recently discovered that even some non-extremal cohomogeneity-two black holes, black rings and microstates are calibrated \cite{Bena:2011ca}. If one could use this to write down a new decomposition of the effective Lagrangian (similar to the one of non-extremal cohomogeneity-one solutions \cite{Gibbons:1982ih,Miller:2006ay,Perz:2008kh,Galli:2011fq}) one would obtain a systematic method to construct new highly-non-trivial and physically-interesting solutions.

\section{The Solution}

We focus on a class of solutions to five-dimensional $N=8$ supergravity that arises as the low-energy limit of a $T^6$ compactification of eleven-dimensional supergravity. 
The spatial part of the five-dimensional spacetime is given by a hyper-K\"ahler space $M_4$, and the warp factor $A$ depends only on the $M_4$ coordinates. The full eleven-dimensional metric is
\begin{equation}
 \diff s_{11}^2 = \; -\e^{-2A}\diff t^2 + \e^{A} \diff s^2(M_4)+ \e^{A} (\diff x_5^2 +\diff x_6^2 + \diff x_7^2+\diff x_8^2) + \e^{-2A}(\diff x_9^2 + \diff x_{10}^2) \,,\label{eq:11d_metric}
\end{equation}
with coordinates $x^5\ldots x^{10}$ on $T^6$. The four-form field strength is 
\eal{
F_4^{\rm mag} \;=\;&~\diff (e^{-3A}) \wedge \diff t \wedge \diff x_9 \wedge \diff x_{10}\\
&+[\Theta_+ - \Theta_-] \wedge \diff x_5\wedge \diff x_8 + [\Theta_+ + \Theta_-]\wedge \diff x_6 \wedge \diff x_7\\
& + \ti \Theta_+\wedge \left(\diff x_6 \wedge \diff x_8-  \diff x_5 \wedge \diff x_7 \right)\label{eq:FourForm}
}
where $\Theta_+,\tilde \Theta_+$ are self-dual two-forms on $M_4$ and $\Theta_-$ is an anti-self-dual one. With hindsight, we focus on a solution whose self-dual forms obey the relation
\begin{equation}\label{eq:hol_2form}
(\Theta_+ + \iu \tilde \Theta_+)\wedge (\Theta_+ + \iu \tilde \Theta_+) = 0 \, ,
\end{equation}
which implies that $\Theta_+ + \iu \tilde \Theta_+$ defines a complex structure on $M_4$ under which it is a holomorphic two-form. As we will see in Section \ref{section:fluxcomp}, this ensures that the solution is supersymmetric.
Finally, the warp factor is determined by 
\begin{equation}\label{eq:WarpFactorEOM}
\Delta_4 e^{3A} = (\Theta_+^2 + \ti\Theta_+^2+\Theta_-^2) + \rho_{M2} \;,
\end{equation}
where $\Delta_4$ is the Laplacian on $M_4$ and $\rho_{M2}$ the M2 brane density.

This solution has the electric charge of a set of M2 branes extended along the $x_9$ and $x_{10}$ directions and smeared on the other compact directions of $T^6$. The magnetic component of the four-form can be thought of as being sourced by four types of M5 branes on the corresponding Poincar\'e dual cycles. We summarize that in Table~\ref{tab:Branes}.

\begin{table}[ht!]
\centering
\begin{tabular}{c|ccccccccc}
&0&9&10&5&6&7&8&$M_4$\\
\hline
M2& $\times$& $\times$ &$\times$&&&&&&\\
M5&$\times$& $\times$&$\times$&$\times$&&&$\times$&$\gamma_1$\\
M5&$\times$& $\times$&$\times$&&$\times$&$\times$&&$\gamma_2$\\
M5&{$\times$}&{$\times$}&{$\times$}&&{$\times$}&&$\times$&$\gamma_3$\\
M5&{$\times$}&{$\times$}&{$\times$}&{$\times$}&&{$\times$}&&$\gamma_4$\\
\end{tabular}
\caption{The brane charges for our configurations along the $T^6$ directions $x_5 \ldots x_{10}$. A brane is localized in directions marked ``$\times$'' and smeared in the other ones.  The M5 branes each wrap a 1-cycle $\gamma_i$ in the hyper-K\"ahler space $M_4$, determined by the (anti)-selfdual fields $\Theta_\pm,\ti \Theta_+$.\label{tab:Branes}}
\end{table}

\subsection{Interpretation as a flux compactification} 
\label{section:fluxcomp}
We  now argue that this solution is a supersymmetric solution of 11-dimensional supergravity. By swapping the roles of $M_4$ and $T^2_{9,10}$ as external and internal spaces, we see that the above solution is actually an eight-dimensional Calabi-Yau `compactification' of M-theory, of the type discussed first in \cite{Becker:1996gj}. The eleven-dimensional spacetime has the form $\calm_{1,10}= \calm_{1,2} \times X_8$, where $X_8 = M_4 \times T^4_{5,6,7,8}$. The metric and the gauge field
preserve three-dimensional Poincar\'e invariance, as can be seen by rewriting \eqref{eq:11d_metric} and \eqref{eq:FourForm} as
\begin{equation} \label{eq:11d_CY4} \begin{aligned}
\diff s_{11}^2 \;=\; & \e^{-2A}(-\diff t^2 + \diff x_{9}^2 + \diff x_{10}^2) + \e^{A} \diff s^2(X_8) \ ,\\
F_4 \; =\; &  \diff (e^{-3A} \operatorname{vol}_3) +\Im\,[(\Theta_+ - \iu\tilde \Theta_+) \wedge \diff z \wedge \diff w + \Theta_-\wedge\diff z \wedge \diff \bar w] \ ,
\end{aligned} 
\end{equation}
where $\operatorname{vol}_3=\diff t \wedge \diff x_{9} \wedge \diff x_{10}$ is the volume form of three-dimensional spacetime and
$A$  only depends on the coordinates of the internal manifold $X_8$. Furthermore, we defined the holomorphic one-forms 
\begin{equation}\label{eq:one-forms_flux}
\diff z = \diff x_5+ \iu \diff x_6\ , \quad \diff w = \diff x_7+ \iu \diff x_8\ .
\end{equation}
The supersymmetry conditions require $\diff s^2(X_8)$ to be a Calabi-Yau metric for $X_8$ and the internal components of $F_4$ to be a primitive $(2,2)$-form. The first two requirements are fulfilled since \eqref{eq:11d_metric} and \eqref{eq:FourForm} give a Calabi-Yau metric
\begin{equation}
\diff s^2(X_8)= \diff s^2(M_4)+\diff z \diff \bar z + \diff w \diff \bar w  \ .
\end{equation}
Since the anti-self-dual two-forms on hyper-K\"ahler manifolds are $(1,1)$, eq. \eqref{eq:11d_CY4} implies that the internal components of $F_4$ indeed make up a primitive $(2,2)$-form if  $(\Theta_+ + \iu \tilde \Theta_+)\wedge \diff z \wedge \diff w$ is the holomorphic four-form of $X_8$ (such that $(\Theta_+ - \iu \tilde \Theta_+)$ is antiholomorphic on $M_4$). This in turn can only be realized if condition \eqref{eq:hol_2form} holds. 
The equation of motion for the gauge field then determines the warp factor in general as 
\begin{equation}\label{eq:WarpFactorEOM2}
\diff \ast_8 \diff \; A = \tfrac 16  F^{\rm mag}_4 \wedge F^{\rm mag}_4 \ ,
\end{equation}
which reduces to \eqref{eq:WarpFactorEOM} when $X_8 = M_4 \times T^4_{5,6,7,8}$.
Note that the described background is dual to a supersymmetric flux background of IIB string theory in the GKP class \cite{Grana:2000jj,Giddings:2001yu}.

\subsection{Relation to five-dimensional STU solutions}
\label{sec:STU}

Finally, we can interpret our supersymmetric solution in eleven-dimensional supergravity compactified on a six-torus ($T^6_{(5,6,7,8,9,10)}$) which descends to five-dimensional $N=8$ supergravity. 
There exists a very large class of solutions to this theory, that fit inside an $N=2$ truncation with two vector multiplets: they describe black rings, black holes as well as microstate solutions that have the same charges as these objects but no horizon.

All supersymmetric solutions of this truncation are known \cite{Gutowski:2004yv,Bena:2004de}, and are given by:
\eal{
\diff s_{11}^2 &=\; -Z^{-2}(\diff t + k)^2 + Z \,\diff s_4^2+Z\sum_{I=1}^3 \frac{\diff s_I^2}{Z_I}\ ,\\
F_4 &=\; \diff A^{(I)}\wedge \omega_I = \sum_{I=1}^3 \left(- \diff \left({\diff t + k \over Z_I}\right) + \Theta^{(I)}\right) \wedge \omega_I\ , \label{eq:ThreeChargeSol}
}
where $Z \equiv (Z_1 Z_2 Z_3)^{1/3}$, $\diff s_I^2$ and $\omega_I$ are respectively a unit metric and a unit volume form on the three $T^2$'s inside $T^6$ and $\diff s_4^2$ is a four-dimensional hyper-K\"{a}hler metric. When this metric has a translational $U(1)$ isometry it becomes a Gibbons-Hawking metric; if one then compactifies along the Gibbons-Hawking fiber, one obtains a solution of the four-dimensional STU model. Note that we work in a convention in which the three curvature two-forms of the  hyper-K\"{a}hler base are self-dual, and hence the $\Theta^{(I)}$ of a supersymmetric solution are anti-self-dual.

The metric and the timelike (electric) components of the four-form of our solution ({\ref{eq:11d_metric},\ref{eq:FourForm}}) are of the form \eqref{eq:ThreeChargeSol} with $Z_1=Z_2=1$ and $k=0$. However, the spacelike (magnetic) four-form field strengths have more components, and only reduce to the $N=2$ truncation above when  $\Theta_+= \ti\Theta_+ =0$. Hence, despite having the right electric charges, the supersymmetric $N=8$ solution we found does not fit into the standard ``STU'' $N=2$ truncation. In the next section we  discuss the supersymmetry of this solution, and how it fits into a larger $N=2$ truncation.

\section{Supersymmetry in $N=8$ and $N=2$}
\label{section:susy}

We have shown already in Section \ref{section:fluxcomp}  that the solution (\ref{eq:11d_metric}, \ref{eq:FourForm}) is a Calabi-Yau four-fold flux background, and hence preserves at least four supercharges \cite{Becker:1996gj}. We first analyze the supersymmetry in detail and then discuss whether the solution and its supercharges fit inside the largest $N=2$ truncation of the $N=8$ theory.

\subsection{$1/8$ BPS solutions in $N=8$ supergravity}\label{section:BPS}

Clearly, the hyper-K\"ahler background breaks half of the supersymmetry, as it admits only a covariant spinor of (say) positive chirality. This corresponds to the projection $\Gamma^{1234} \eta = - \eta$, where $\eta$ is a spinor on the internal eight-dimensional manifold. Furthermore, the flux $F_4$ breaks more supersymmetry. Its electric component (corresponding to an M2-brane charge along the $9,10$ directions) breaks another half of supersymmetry, by the projection $\Gamma^{12345678} \eta = \eta$. 

To understand how the magnetic components of $F_4$ affect the supersymmetry, it is best to choose an appropriate vierbein $e^i, i=1,\dots,4,$ on the hyper-K\"ahler space $M_4$, such that  \eqref{eq:hol_2form} is fulfilled and we can identify the self-dual two-forms of \eqref{eq:FourForm} as
\eal{
\Theta_+ &=\; \theta_+ (e^1 \wedge e^3 + e^4 \wedge e^2)\ ,\\
\tilde \Theta_+ &=\; \theta_+(e^1 \wedge e^4 + e^2\wedge e^3) \ .\label{eq:Theta_+}
}
The supersymmetry conditions $\slashed F \eta = 0$ and $\slashed F_m \eta = 0$ \cite{Becker:1996gj} will now contain an additional projector, which further halves the amount of supersymmetry. More precisely:
\eal{
0=\tfrac 1 {4!}F_{ijkl} \Gamma^{ijkl} \eta \;=\;& \tfrac14[(\Theta_+)_{ij}\, \Gamma^{ij58} +(\tilde\Theta_+)_{ij}\,\Gamma^{ij68} ](1 - \Gamma^{5678})(1 -\Gamma^{1234})\eta\\
&- \tfrac14(\Theta_-)_{ij}\, \Gamma^{ij58} (1 + \Gamma^{5678})(1 +\Gamma^{1234})\eta\ ,
}
where we have inserted the projectors $\tfrac12(1 \pm \Gamma^{1234})$ by making use of the (anti-)self-duality of $\Theta_\mp$. 

The term containing the anti-self-dual flux $\Theta_-$ vanishes on the Killing spinors annihilated by the two earlier projectors $\tfrac12 (1+\Gamma^{1234})$ and $\tfrac12 (1-\Gamma^{12345678}) $, and this agrees with the known structure of BPS three-charge solutions, in which turning on an anti-self-dual field strength on the base does not affect supersymmetry. 

For arbitrary self-dual forms $\Theta_+, \ti \Theta_+$, the first line is not zero and supersymmetry is broken. However, for the specific choice \eqref{eq:Theta_+} this term contains a new projector:
\begin{equation}
0 =  \, 2 \theta_+ \Gamma^{1358}(1 + \Gamma^{3456})\eta \ ,
\end{equation}
which is compatible with the first two.  More generally, under the condition \eqref{eq:hol_2form} we always find such a projector and the solution has four supercharges.

It is not hard to see that the equations $\slashed F_m \eta = 0$ do not impose any extra conditions on the remaining Killing spinors, essentially because the flux pieces that are self-dual on the hyper-K\"ahler manifold always combine into the projector $\tfrac12(1 + \Gamma^{3456})$, while the anti-self-dual components always give either $\tfrac12(1 +\Gamma^{1234})$ or $\tfrac12(1 + \Gamma^{5678})$, depending on the index $m$.
Therefore, the solution is $1/8$ BPS, and its 4 Killing spinors are annihilated by the projectors:
\begin{equation}\label{eq:projectors}
\tfrac12(1+\Gamma_{1234})\ ,\  \tfrac12(1 + \Gamma_{3456}) \ {\rm and }\ \tfrac12(1+\Gamma_{5678}) \ .
\end{equation}

\subsection{A puzzle} \label{section:puzzle}

The $1/8$ BPS solution we gave in (\ref{eq:11d_metric},\ref{eq:FourForm}) has not been found in the literature. Moreover, its magnetic field strength \eqref{eq:FourForm} has both self-dual and anti-self-dual components on the hyper-K\"ahler space. This is surprising since all 1/2 BPS solutions in $N=2$ supergravity in five dimensions have only anti-self-dual fluxes on the hyper-K\"ahler space, as shown in \cite{Gauntlett:2002nw,Gutowski:2004yv}. This indicates that our solution cannot be a 1/2 BPS solution of $N=2$ supergravity. In the following we want to discuss what happens to the $1/8$ BPS solution (\ref{eq:11d_metric},\ref{eq:FourForm}) when mapped to the maximal $N=2$ truncation of $N=8$ supergravity. 

\subsection{$N=2$ truncations and supersymmetry} \label{section:N=2}

In order to find a supergravity with eight supercharges in five dimensions, we have to perform a truncation of $N=8$ supergravity. The field content of these truncated theories (also called `magical supergravities') has been discussed for instance in \cite{Gunaydin:1983rk,Gunaydin:1983bi}. The $N=2$ truncation with the maximal field content (and only vector multiplets) is the magical supergravity related to the Jordan algebra over the quaternions and it admits the global symmetry group $SU^*(6)$. It has the same bosonic field content as five-dimensional $N=6$ supergravity. As we show in a more detailed work \cite{Bena:2011aa}, the projection to this $N=2$  supergravity in five dimensions corresponds to fixing a complex structure $I$ on $T^6$ and projecting out some representations of the related $\SLn 3 C$. The surviving vector fields of the $N=2$ projection contain all gauge fields coming from the eleven-dimensional three-form potential with two legs on $T^6$ that are $(1,1)$ with respect to $I$. Note that $I$ does not have to be related to the complex structure under which $\diff z$ and $\diff w$ are holomorphic, as long as the metric given in \eqref{eq:11d_metric} respects it.  If we choose a complex structure $I$ on $T^6$ such that 
\begin{equation}\label{new_complex_structure}
\diff z^1 =  \diff x^8 + \iu \diff x^5 \ , \
\diff z^2 =  \diff x^6 + \iu \diff x^7 \ {\rm and} \
\diff z^3 =  \diff x^9 + \iu \diff x^{10} 
\end{equation}
are holomorphic one-forms under $I$, then the flux given in \eqref{eq:FourForm} is $(1,1)$ on $T^6$, and we see that our solution indeed gives a solution to $N=2$ supergravity.

Now let us understand the amount of supersymmetry of the solution in $N=2$ supergravity. The complex structure above is different from the complex structure chosen in \eqref{eq:one-forms_flux}, and under the new complex structure the flux $F_4$ \eqref{eq:11d_CY4} has a piece that is $(3,1)\oplus (1,3)$ and therefore the configuration is  not supersymmetric in $N=2$ supergravity. 
More precisely, the projection to $N=2$ breaks the $N=8$ R-symmetry group $USp(8)$ to $USp(6) \times SU(2)$, where the latter factor is the R-symmetry of the $N=2$ theory. The action of $USp (6)$ on the spinors defines the projection to $N=2$. The generator 
\begin{equation}
 C \equiv \tfrac12 (\Gamma^{85} - \Gamma^{67})
\end{equation}
commutes with the complex structure $I$, the Cartan generator of $SU(2)$,  and hence is a generator of $USp(6)$.  In particular, the requirement $C\eta = 0$ implies
\begin{equation}
\tfrac12 (1-\Gamma^{5678}) \eta = 0 \ . 
\end{equation}
This projects out all four Killing spinors of the $1/8$ BPS solution, cf.\ \eqref{eq:projectors}. 
Hence, when we projected to the $N=2$ $SU^*(6)$ supergravity, we projected out all supercharges which remain unbroken in the solution (\ref{eq:11d_metric}, \ref{eq:FourForm}). Therefore, the solution is non-BPS in $N=2$ supergravity.

\section*{Acknowledgments}
We are very grateful for useful discussions with S.\ Ferrara, S.\ Giusto, M.\ Gra\~na, M.\ Gunaydin and M.\ Shigemori.
We thank the Aspen Center for Physics and the Centro de Ciencias de Benasque Pedro
Pascual for hospitality while this work was completed. This work was supported in part by the ANR grant 08-JCJC-0001-0,
by the ERC Starting Independent Researcher Grant 240210 - String-QCD-BH as well as by the
Aspen Center for Physics NSF Grant 1066293.

\bibliographystyle{toine}
\bibliography{FuzzComp}

\providecommand{\href}[2]{#2}\begingroup\raggedright\begin{thebibliography}{10}

\bibitem{Grana:2000jj}
M.~Grana and J.~Polchinski, \emph{{Supersymmetric three form flux perturbations
  on AdS(5)}}, Phys.Rev. {\bf D63} (2001) 026001,
\href{http://www.arXiv.org/abs/hep-th/0009211}{{\tt hep-th/0009211}}

\bibitem{Giddings:2001yu}
S.~B. Giddings, S.~Kachru  and J.~Polchinski, \emph{{Hierarchies from fluxes in
  string compactifications}}, Phys.Rev. {\bf D66} (2002) 106006,
\href{http://www.arXiv.org/abs/hep-th/0105097}{{\tt hep-th/0105097}}

\bibitem{Grana:2004bg}
M.~Grana, R.~Minasian, M.~Petrini  and A.~Tomasiello, \emph{{Supersymmetric
  backgrounds from generalized Calabi-Yau manifolds}}, JHEP {\bf 0408} (2004)
  046,
\href{http://www.arXiv.org/abs/hep-th/0406137}{{\tt hep-th/0406137}}

\bibitem{Gauntlett:2002nw}
J.~P. Gauntlett, J.~B. Gutowski, C.~M. Hull, S.~Pakis  and H.~S. Reall,
  \emph{{All supersymmetric solutions of minimal supergravity in five
  dimensions}}, Class. Quant. Grav. {\bf 20} (2003) 4587--4634,
\href{http://www.arXiv.org/abs/hep-th/0209114}{{\tt hep-th/0209114}}

\bibitem{Gutowski:2004yv}
J.~B. Gutowski and H.~S. Reall, \emph{{General supersymmetric AdS(5) black
  holes}}, JHEP {\bf 04} (2004) 048,
\href{http://www.arXiv.org/abs/hep-th/0401129}{{\tt hep-th/0401129}}

\bibitem{Bena:2004de}
I.~Bena and N.~P. Warner, \emph{{One ring to rule them all ... and in the
  darkness bind them?}}, Adv. Theor. Math. Phys. {\bf 9} (2005) 667--701,
\href{http://www.arXiv.org/abs/hep-th/0408106}{{\tt hep-th/0408106}}

\bibitem{Goldstein:2008fq}
K.~Goldstein and S.~Katmadas, \emph{{Almost BPS black holes}}, JHEP {\bf 0905}
  (2009) 058,
\href{http://www.arXiv.org/abs/0812.4183}{{\tt 0812.4183}}

\bibitem{Ferrara:1997tw}
S.~Ferrara, G.~W. Gibbons  and R.~Kallosh, \emph{Black holes and critical
  points in moduli space}, Nucl. Phys. {\bf B500} (1997) 75--93,
\href{http://www.arXiv.org/abs/hep-th/9702103}{{\tt hep-th/9702103}}

\bibitem{Denef:2000nb}
F.~Denef, \emph{{Supergravity flows and D-brane stability}}, JHEP {\bf 0008}
  (2000) 050,
\href{http://www.arXiv.org/abs/hep-th/0005049}{{\tt hep-th/0005049}}

\bibitem{Ceresole:2007wx}
A.~Ceresole and G.~Dall'Agata, \emph{Flow equations for non-BPS extremal black
  holes}, JHEP {\bf 03} (2007) 110,
\href{http://www.arXiv.org/abs/hep-th/0702088}{{\tt hep-th/0702088}}

\bibitem{Andrianopoli:2007gt}
L.~Andrianopoli, R.~D'Auria, E.~Orazi  and M.~Trigiante, \emph{First Order
  Description of Black Holes in Moduli Space},
\href{http://www.arXiv.org/abs/arXiv:0706.0712 [hep-th]}{{\tt arXiv:0706.0712
  [hep-th]}}

\bibitem{LopesCardoso:2007ky}
G.~L. Cardoso, A.~Ceresole, G.~Dall'Agata, J.~M. Oberreuter  and J.~Perz,
  \emph{First-order flow equations for extremal black holes in very special
  geometry}, JHEP {\bf 10} (2007) 063,
\href{http://www.arXiv.org/abs/0706.3373}{{\tt 0706.3373}}

\bibitem{Lust:2008zd}
D.~Lust, F.~Marchesano, L.~Martucci  and D.~Tsimpis, \emph{{Generalized
  non-supersymmetric flux vacua}}, JHEP {\bf 0811} (2008) 021,
\href{http://www.arXiv.org/abs/0807.4540}{{\tt 0807.4540}}

\bibitem{Held:2010az}
J.~Held, D.~Lust, F.~Marchesano  and L.~Martucci, \emph{{DWSB in heterotic flux
  compactifications}}, JHEP {\bf 1006} (2010) 090,
\href{http://www.arXiv.org/abs/1004.0867}{{\tt 1004.0867}}

\bibitem{Becker:1996gj}
K.~Becker and M.~Becker, \emph{{M theory on eight manifolds}}, Nucl.Phys. {\bf
  B477} (1996) 155--167,
\href{http://www.arXiv.org/abs/hep-th/9605053}{{\tt hep-th/9605053}}

\bibitem{Bena:2011aa}
I.~Bena, M.~Grana, H.~Triendl  and B.~Vercnocke,
\emph{{To Appear}},

\bibitem{Ferrara:2007pc}
S.~Ferrara and A.~Marrani, \emph{{N=8 non-BPS Attractors, Fixed Scalars and
  Magic Supergravities}}, Nucl.Phys. {\bf B788} (2008) 63--88,
\href{http://www.arXiv.org/abs/0705.3866}{{\tt 0705.3866}}

\bibitem{Khuri:1995xq}
R.~R. Khuri and T.~Ortin, \emph{{A Nonsupersymmetric dyonic extreme
  Reissner-Nordstrom black hole}}, Phys.Lett. {\bf B373} (1996) 56--60,
\href{http://www.arXiv.org/abs/hep-th/9512178}{{\tt hep-th/9512178}}

\bibitem{Ferrara:2006yb}
S.~Ferrara, E.~G. Gimon  and R.~Kallosh, \emph{{Magic supergravities, N= 8 and
  black hole composites}}, Phys.Rev. {\bf D74} (2006) 125018,
\href{http://www.arXiv.org/abs/hep-th/0606211}{{\tt hep-th/0606211}}

\bibitem{Bossard:2010mv}
G.~Bossard, \emph{{1/8 BPS Black Hole Composites}},
\href{http://www.arXiv.org/abs/1001.3157}{{\tt 1001.3157}}

\bibitem{Bates:2003vx}
B.~Bates and F.~Denef, \emph{Exact solutions for supersymmetric stationary
  black hole composites},
\href{http://www.arXiv.org/abs/hep-th/0304094}{{\tt hep-th/0304094}}

\bibitem{Elvang:2004rt}
H.~Elvang, R.~Emparan, D.~Mateos  and H.~S. Reall, \emph{{A supersymmetric
  black ring}}, Phys. Rev. Lett. {\bf 93} (2004) 211302,
\href{http://www.arXiv.org/abs/hep-th/0407065}{{\tt hep-th/0407065}}

\bibitem{Elvang:2004ds}
H.~Elvang, R.~Emparan, D.~Mateos  and H.~S. Reall, \emph{{Supersymmetric black
  rings and three-charge supertubes}}, Phys. Rev. {\bf D71} (2005) 024033,
\href{http://www.arXiv.org/abs/hep-th/0408120}{{\tt hep-th/0408120}}

\bibitem{Gauntlett:2004qy}
J.~P. Gauntlett and J.~B. Gutowski, \emph{General concentric black rings},
  Phys. Rev. {\bf D71} (2005) 045002,
\href{http://www.arXiv.org/abs/hep-th/0408122}{{\tt hep-th/0408122}}

\bibitem{Bena:2011zw}
I.~Bena, B.~D. Chowdhury, J.~de~Boer, S.~El-Showk  and M.~Shigemori,
  \emph{{Moulting Black Holes}}, \href{http://www.arXiv.org/abs/1108.0411}{{\tt
  1108.0411}},
* Temporary entry *

\bibitem{Bena:2009fi}
I.~Bena, S.~Giusto, C.~Ruef  and N.~P. Warner, \emph{{Supergravity Solutions
  from Floating Branes}}, JHEP {\bf 1003} (2010) 047,
\href{http://www.arXiv.org/abs/0910.1860}{{\tt 0910.1860}}

\bibitem{Bena:2011ca}
I.~Bena, C.~Ruef  and N.~P. Warner, \emph{{Imaginary Soaring Branes: A Hidden
  Feature of Non-Extremal Solutions}},
\href{http://www.arXiv.org/abs/1105.6255}{{\tt 1105.6255}}

\bibitem{Gibbons:1982ih}
G.~Gibbons, \emph{{Antigravitating Black Hole Solitons with Scalar Hair in N=4
  Supergravity}}, Nucl.Phys. {\bf B207} (1982)
337--349

\bibitem{Miller:2006ay}
C.~M. Miller, K.~Schalm  and E.~J. Weinberg, \emph{{Nonextremal black holes are
  BPS}}, Phys.Rev. {\bf D76} (2007) 044001,
\href{http://www.arXiv.org/abs/hep-th/0612308}{{\tt hep-th/0612308}}

\bibitem{Perz:2008kh}
J.~Perz, P.~Smyth, T.~Van~Riet  and B.~Vercnocke, \emph{{First-order flow
  equations for extremal and non-extremal black holes}}, JHEP {\bf 03} (2009)
  150,
\href{http://www.arXiv.org/abs/0810.1528}{{\tt 0810.1528}}

\bibitem{Galli:2011fq}
P.~Galli, T.~Ortin, J.~Perz  and C.~S. Shahbazi, \emph{{Non-extremal black
  holes of N=2, d=4 supergravity}}, JHEP {\bf 1107} (2011) 041,
  \href{http://www.arXiv.org/abs/1105.3311}{{\tt 1105.3311}},
* Temporary entry *

\bibitem{Gunaydin:1983rk}
M.~Gunaydin, G.~Sierra  and P.~K. Townsend, \emph{{Exceptional Supergravity
  Theories and the MAGIC Square}}, Phys. Lett. {\bf B133} (1983)
72

\bibitem{Gunaydin:1983bi}
M.~Gunaydin, G.~Sierra  and P.~K. Townsend, \emph{{The Geometry of N=2
  Maxwell-Einstein Supergravity and Jordan Algebras}}, Nucl. Phys. {\bf B242}
  (1984)
244

\end{thebibliography}\endgroup

\end{document}